\documentclass{80SA}
\pdfoutput=1
\usepackage{times}
\usepackage[T1]{fontenc}
\usepackage[latin1]{inputenc}
\usepackage{graphicx}

\newcommand{\YRS}{YbRh$_2$Si$_2$}
\newcommand{\YCxS}{Yb(Rh$_{1-x}$Co$_x$)$_2$Si$_2$}
\newcommand{\RH}{\ensuremath{R_{\text H}}}
\newcommand{\TN}{\ensuremath{T_{\text N}}}
\newcommand{\TL}{\ensuremath{T_{\text L}}}
\newcommand{\BN}{\ensuremath{B_{\text N}}}
\newcommand{\Tstar}{\ensuremath{T^{\star}}}
\newcommand{\Bstar}{\ensuremath{B^{\star}}}

\newcommand{\sven}[1]{%
{#1}}

\title{Break up of heavy fermions at an antiferromagnetic instability}

\author{S. Friedemann$^1$\thanks{E-mail address: sf425@cam.ac.uk \newline Present address: University of Cambridge, United Kingdom.}, S. Wirth$^1$, S. Kirchner$^{1,2}$, Q. Si$^{3}$,  S. Hartmann$^1$, C. Krellner$^1$, C. Geibel$^1$, T. Westerkamp$^1$, M. Brando$^1$ and F. Steglich$^1$}
\inst{$^1$Max Planck Institute for Chemical Physics of Solids, 01187 Dresden, Germany; 
$^2$Max Planck Institute for Physics of Complex Systems, 01187 Dresden, Germany; 
$^{3}$Department of Physics and Astronomy, Rice University, Houston, TX 77005, USA}

\abst{We present results of high-resolution, low-temperature measurements of the Hall coefficient, thermopower, and specific heat
on stoichiometric \YRS. They support earlier conclusions of an electronic (Kondo-breakdown) quantum critical point concurring with a field induced antiferromagnetic one. We also discuss the detachment of the two instabilities under chemical pressure. Volume compression/expansion (via substituting Rh by Co/Ir) results in a stabilization/weakening of magnetic order. Moderate Ir substitution leads to a non-Fermi-liquid phase, in which the magnetic moments are neither ordered nor screened by the Kondo effect. The so-derived zero-temperature global phase diagram promises future studies to explore the nature of the Kondo breakdown quantum critical point without any interfering magnetism.
}

\kword{quantum criticality, heavy fermion, \YRS, Kondo breakdown
}

\begin{document}
\maketitle

\section{Different types of quantum critical points}
Quantum phase transitions in matter arise due to competing interactions, which result in competing ground-state properties. When a quantum phase transition is continuous it marks a quantum critical point (QCP). In the last decade, antiferromagnetic (AF) heavy-fermion metals turned out to be model systems to study quantum criticality. In these systems, QCPs are caused by the competition between the local Kondo and the non-local Ruderman-Kittel-Kasuya-Yoshida (RKKY) interaction. Studies on the interplay between these two phenomena have revealed different types of QCPs \cite{Gegenwart08}. In the itinerant, spin-density-wave (SDW) scenario the heavy fermions 
keep their integrity at the QCP. In such a case the QCP can be treated as a continuous classical phase transition in an effective dimension $d+z$, where $d$ is the spatial dimensionality and $z$, the dynamic exponent defined via $\xi_t\sim\xi_r^z$, describes the number of additional spatial dimensions that the time dimension corresponds to. $\xi_r$ and $\xi_t$ denote the correlation length and correlation time which both diverge at a QCP \cite{Hertz76,Moriya85,Millis93}. Several heavy-fermion compounds, \textit{e.~g.}, CeCu$_{2}$Si$_{2}$ \cite{Gegenwart98}, CeNi$_{2}$Ge$_{2}$ \cite{Gegenwart99} and Ce$_{1-x}$La$_{x}$Ru$_{2}$Si$_{2}$ \cite{Knafo09} were found to exhibit this type of itinerant AF QCP.

However, in a few heavy-fermion metals the AF instability appears to be accompanied by a breakdown of the Kondo effect \cite{Si01,Coleman01,Senthil04}, which is sometimes called a zero-temperature, $f$-electron selective Mott transition \cite{Paul07}. CeCu$_{5.9}$Au$_{0.1}$ \cite{Loehneysen94,Schroeder00} and YbRh$_{2}$Si$_{2}$ \cite{Trovarelli00} are prototypical examples for such an unconventional quantum critical behavior. Here, we discuss some of the unique properties of YbRh$_{2}$Si$_{2}$ in the vicinity of its AF QCP. In addition, we show how the AF and Kondo breakdown QCPs can be detached under chemical pressure.

\section{Concurrence of AF and Kondo breakdown QCPs in YbRh$_{2}$Si$_{2}$}
YbRh$_{2}$Si$_{2}$ crystallizes in the body-centered tetragonal ThCr$_{2}$Si$_{2}$ structure. 
The lowest-lying of the crystal-field Kramer's doublets is well separated from the excited states, so that at $T\leq$ 1~K, YbRh$_{2}$Si$_{2}$ is an effective spin $1/2$ system. It exhibits very weak AF order below $\TN = 70$\,mK, which is continuously suppressed by 
a small critical magnetic field 
\BN\ (60~mT for field applied perpendicular to the crystallographic $c$ axis, 660~mT  for $B \parallel c$) 
\cite{Custers03} accessing the QCP.
On either side of the QCP, 
 heavy Fermi-liquid (FL) behavior was found such as a quadratic form of the resistivity $\rho(T) = \rho_0 + A T^2$ where $\rho_0$ denotes the residual resistivity\cite{Gegenwart02}.
Approaching \BN\ leads to a suppression of the FL behavior and a divergence of $A \propto (B-\BN)^{-1}$ indicating a critical slowing down of the heavy fermions at the QCP. Right at \BN, pronounced non-Fermi-liquid (NFL) effects are observed down to the lowest accessible temperatures ($\approx$ 10~mK), notably a linear $T$ dependence of the electrical resistivity \cite{Gegenwart08a}. 
We note that the quadratic form at \BN\ reported in Ref.~\cite{Knebel06} might easily arise from either incorrect fine tuning of $B$ or heating effects.
The NFL form of $\rho(T)$ is accompanied by a logarithmic divergence of the Sommerfeld coefficient $\gamma$ of the electronic specific heat, $C_{\text{ el}}(T)=\gamma T$, between 0.3 and 10~K. Below 0.3~K, $\gamma(T)$ diverges stronger than logarithmically ($\propto T^{-1/3}$) \cite{Custers03}.

The two FL phases adjacent to the QCP appear to posses different Fermi surfaces as inferred from a variety of electronic transport measurements closely related to the Fermi surface properties.
Most important evidence stems from Hall effect measurements in the so called crossed-field geometry. Here, two perpendicular magnets are used to disentangle the two tasks of the magnetic field: One field, $B_1$, generates the Hall response, and a second field, $B_2$, tunes the ground state of the sample. The power of the crossed-field setup lies in the ability to extract the initial-slope Hall coefficient as a linear response to $B_1$ despite measuring at a finite tuning field $B_2$. 

\begin{figure}
	\begin{center}
		\includegraphics[width=\columnwidth]{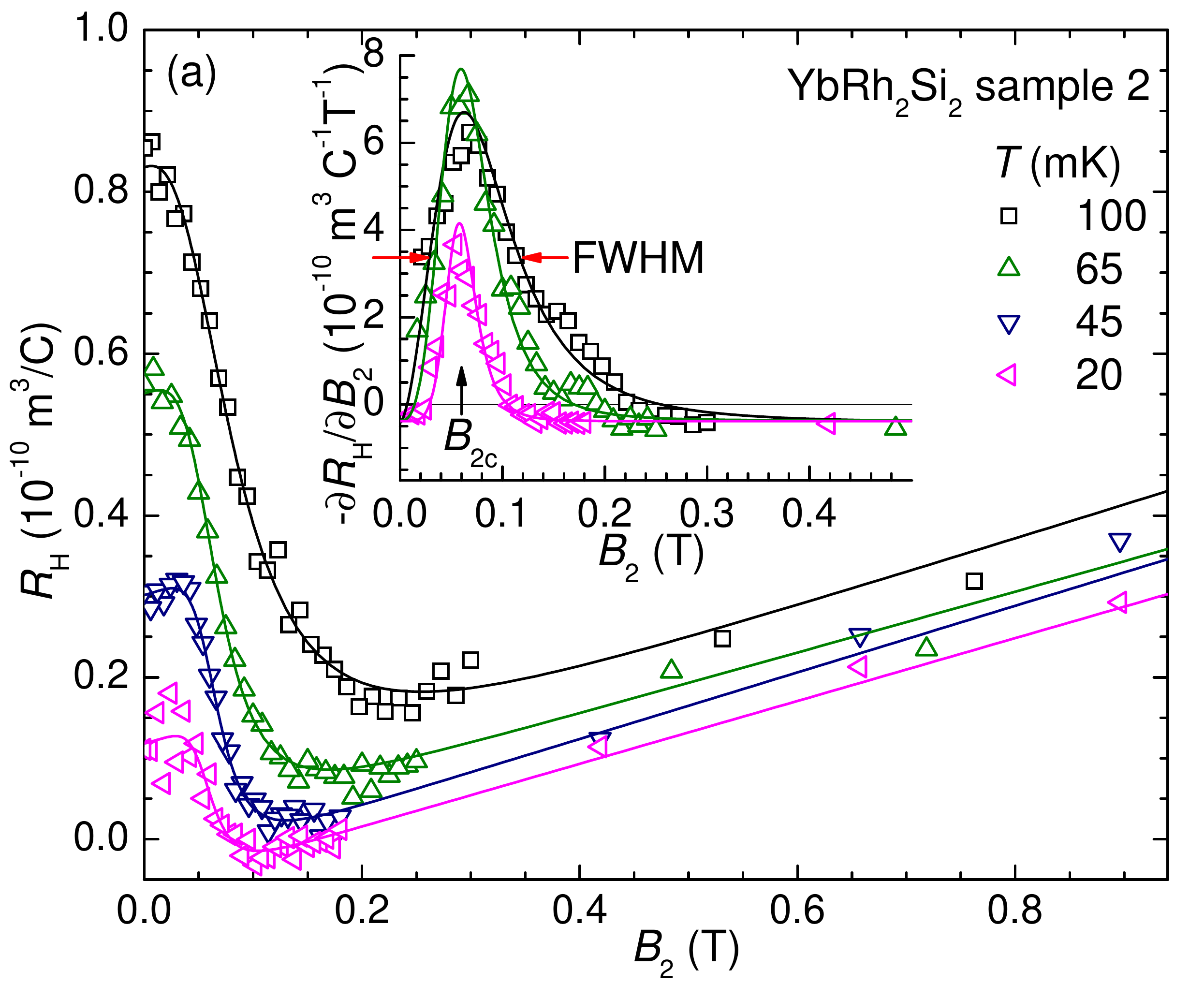}\\
		\includegraphics[width=\columnwidth]{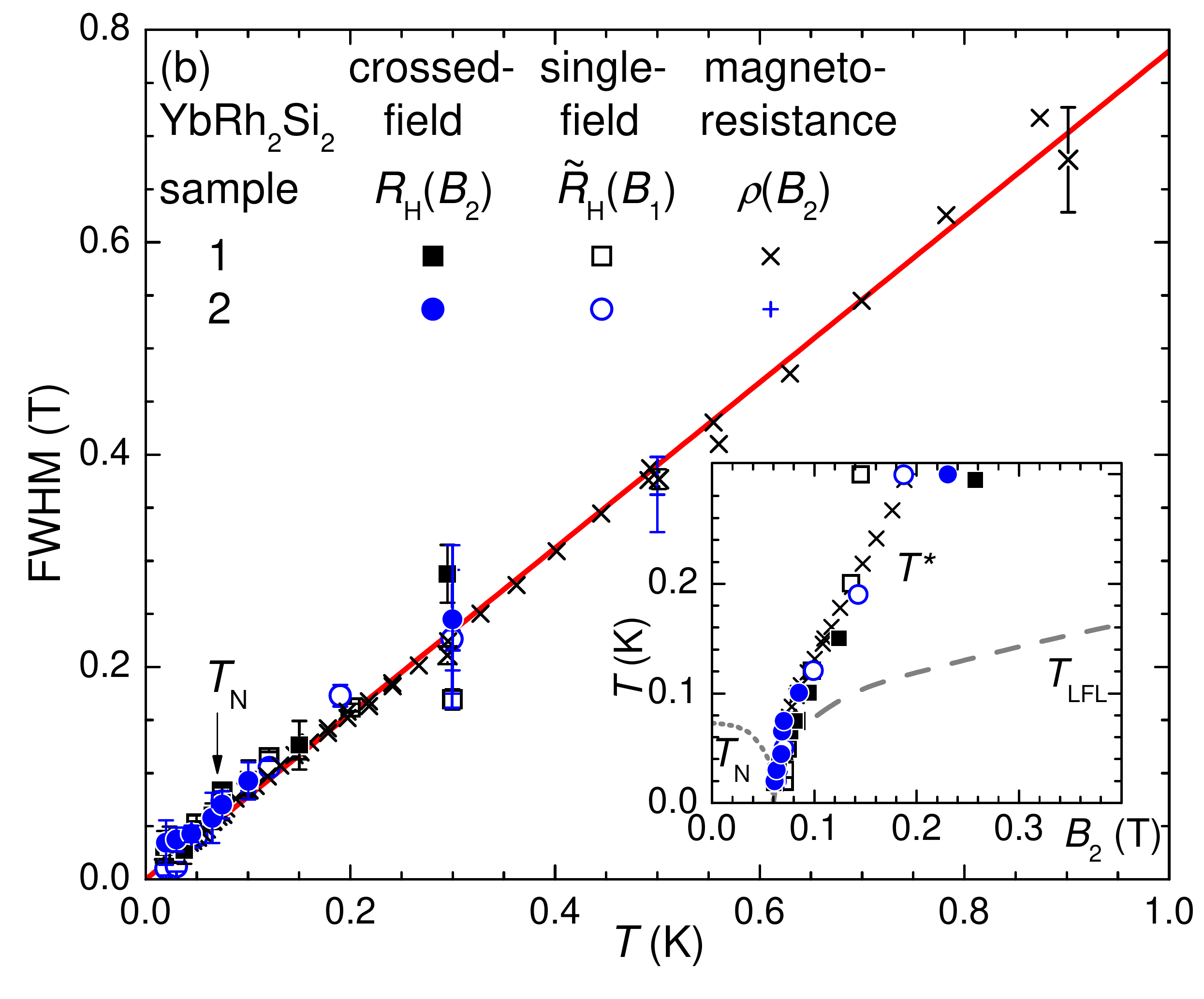}
	\end{center}
	\caption{Crossover in the Hall coefficient of \YRS. (a) The dependence of the Hall coefficient \RH\ on the tuning field $B_2$ applied within the crystallographic $ab$ plane was measured in crossed-field geometry \cite{Friedemann10}. Solid lines correspond to fits of a crossover function superposed to a linear background \cite{Friedemann10}. Inset shows the derivative $- \partial \RH / \partial B_2$. Vertical arrow denotes the critical field $\BN$ ($\bot c$) whereas horizontal arrows illustrate the determination of the full width at half maximum (FWHM) of the crossover for one temperature. (b) Temperature dependence of the FWHM of the crossover in crossed-field ($\RH(B_2)$) and single-field ($\tilde{R}_{\text H}(B_1)$) Hall effect and magnetoresistivity ($\rho (B_2)$). The solid line marks a linear fit to all datasets intersecting the origin within experimental accuracy. Arrow indicates the Néel temperature. Inset in (b) depicts the position of the crossover in the magnetic field-temperature phase diagram \cite{Friedemann10}. The datasets from the different experiments largely agree with each other.
	 Dotted and dashed line reflect the boundary of the antiferromagnetically ordered ground state and of the field-induced Fermi-liquid state, respectively.}
	\label{fig:RHvsB2}
\end{figure}

Isotherms of the field dependent Hall coefficient $\RH(B_2)$ depicted in Fig.~\ref{fig:RHvsB2}(a) show a crossover as \YRS\ is tuned across its QCP \cite{Paschen04,Friedemann10}. The crossover is situated on top of a background contribution. 
The two contributions are further illustrated in the inset of Fig.~\ref{fig:RHvsB2}(a) plotting the derivative $- \partial \RH / \partial B_2$. Here, the crossover corresponds to the peak in proximity to the critical field whereas the linear background is seen as an offset. Despite strong sample dependences observed in  the low-$T$ Hall coefficient, the crossover appears to be robust. Rather, the sample dependences are associated with the background contribution \cite{Friedemann10}. In fact, the considered single crystals which span the hole range of sample dependences \cite{Friedemann10a} show that the crossover persists in its extrapolation to zero temperature \cite{Friedemann10b}. 
The change from a positive to a negative value of \RH\ observed
in this nearly compensated metal at $T=20$\,mK
was shown to be consistent with results from (renormalized) band structure calculations \cite{Friedemann10a}.
As the temperature is lowered, the crossover shifts to lower fields as illustrated in the phase diagram in the inset of Fig.~\ref{fig:RHvsB2}(b) plotting the crossover field $B_0$ (see Ref.~\cite{Friedemann10} for the the fitting procedure used to extract $B_0$). In the extrapolation to zero temperature $B_0$ converges to the QCP thus showing that the quantum criticality is the origin of the crossover in $\RH(B_2)$.

The crossover sharpens as the temperature is lowered. The full width at half maximum (FWHM) of the peak in $- \partial \RH / \partial B_2$ 
is displayed in Fig.~\ref{fig:RHvsB2}(b). Importantly, the FWHM is proportional to temperature, \textit{i.e.}, extrapolates to zero for $T\to 0$. This finding reflects the unconventional nature of the QCP in \YRS. On the one hand, the vanishing width of the Hall crossover at zero temperature implies a discontinuity of the Hall coefficient and, hence, a discontinuous evolution of the Fermi surface at the QCP \cite{Friedemann10b}. Such a Fermi surface reconstruction is incompatible with the smooth crossover predicted for an SDW QCP. Rather, the Fermi surface reconstruction points towards a disintegration of the quasiparticles due to the breakdown of the Kondo effect. On the other hand, the FWHM being proportional to temperature is an indicator for energy over temperature, $E/T$, scaling of the single electron excitations  as the width of the Fermi surface crossover and thus also of the Hall crossover can be associated with the relaxation rate $\Gamma$ of the single electron excitations \cite{Friedemann10}. Such an $E/T$ scaling is fundamentally inconsistent with the $E/T^{3/2}$ scaling and the concomitant superlinear temperature dependence of $\Gamma$ and the FWHM  expected for a SDW QCP in three dimensions (3D).
Consequently, both the fundamental signatures of a Kondo breakdown QCP, the Fermi surface collapse and the unconventional scaling behavior are seen in \YRS. 

The crossover in $\RH(B_2)$ is accompanied by signatures in thermodynamic and transport properties which establish a new energy scale $\Tstar(B)$ in the phase diagram \cite{Gegenwart07}. This energy scale is associated with the Kondo breakdown and for the case of stoichiometric \YRS\ at ambient pressure, it converges together with the magnetic phase boundary and the crossover to the FL regime towards the QCP (cf.\ inset of Fig.~\ref{fig:RHvsB2}(b)).

\begin{figure}
	\begin{center}
		\includegraphics[width=\columnwidth]{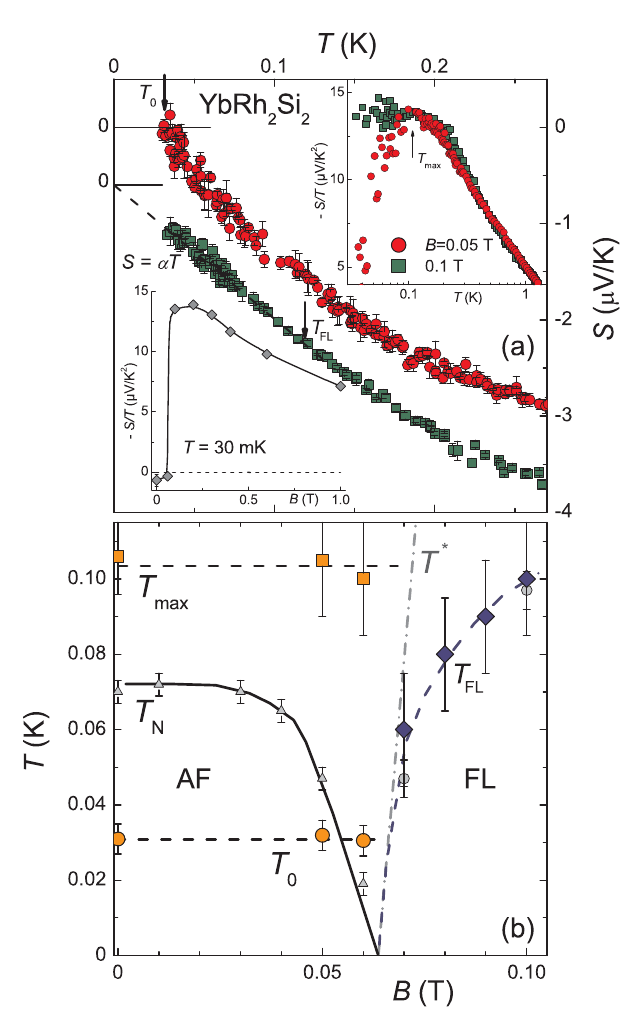}
	\end{center}
	\caption{(a) Low-$T$ thermopower $S(T)$ of \YRS\ at $B$ = 0.05 and 0.1~T applied within the $ab$ plane. Data sets are separated by $\Delta S= -0.6\,\mu$V/K for clarity, solid lines indicate $S=0$ for each curve. Arrow indicates sign change at $T_0\approx$ 30~mK for $B = 0.05$\,T curve. Dashed line represents a linear fit, $S=\alpha_0 T$, to  $B$ = 0.1~T data below NFL - FL crossover temperature $T_{\text{ FL}}$ (cf.~arrow). 
	Upper inset: Thermopower divided by temperature, $-S(T)/T$. 
	Lower inset: Thermopower isotherm, $-S(B)/T$. 
	(b) $T-B$ - phase diagram of \YRS\ around $\BN$. Temperatures $T_0$ and $T_{\text{ max}}$ of the sign change and of the maximum in $-S(T)/T$ are included, respectively. Néel temperature $T_{\text N}$ was derived from $\rho(T)$ results on the same sample (gray triangles); NFL - FL crossover temperature $T_{\text{ FL}}$ either from $\rho(T)$ (onset of $\Delta\rho\propto T^2$, gray circles) or from $S(T)$ measurements (onset of $S\propto T$, diamonds). $T_{\text N}(B)$ and $T_{\text{ FL}}(B)$ extrapolate to a critical field $\BN\approx$ 64~mT, lines are guides to the eye. The crossover line $\Tstar(B)$ was taken from \cite{Gegenwart07} and was scaled to $\BN$ of the particular sample.}
	\label{fig:thermopower}
\end{figure}

The change of the low-$T$ Hall coefficient from hole- to electron-dominated transport (upon increasing $B$ through \BN)
 is further confirmed by low-temperature thermopower $S$ results \cite{Hartmann10}.
For $B<\BN$ a change from negative to positive $S$ occurs when cooling to below $T_0$ = 30\,mK, whereas for $B>\BN$ the results indicate  negative values down to $T=0$. 
In the latter regime the thermopower is proportional to temperature,
$S(T)=\alpha_0 T$,
which is typical for a FL. 
In the upper inset of Fig.~\ref{fig:thermopower}(a), the saturation of $S(T)/T$ to a constant value below $T_{\text{ FL}}$ reflects the FL character of the ground state on the high-field side of the QCP. 
At all fields $B<\BN$, $-S(T)/T$ 
exhibits a pronounced maximum  succeeded by a  drastic drop at lower temperatures.
This anomaly at $T_{\text{ max}}\approx 100$\,mK as well as the sign change at $T_0$ are unrelated to the magnetic phase boundary as illustrated in Fig.~\ref{fig:thermopower}(b). Rather, $T_0$ and $T_{\text{ max}}$ are constant up to $B\approx \BN$  where they vanish discontinuously. This is to be contrasted with the continuous nature of the magnetic transition, which is smoothly suppressed at $\BN$.

In addition, the magnetic field dependence of the thermopower exhibits a drastic change upon crossing the $\Tstar(B)$ line. The maximum observed in the isothermal thermopower coefficient, $-S(B)/T$, sharpens and shifts to lower fields upon cooling. The inflection point at the low-field side of the anomaly coincides well with the energy scale $\Tstar(B)$, cf. lower inset in Fig.~\ref{fig:thermopower}(a) \cite{Hartmann10}. In the $T=0$ limit, the field-dependence of $S/T$ is expected to develop into a step-like function, \textit{i.~e.} a discontinuity to occur exactly at the QCP, indicating an abrupt change of the Fermi surface at the QCP. 

The discontinuities in $S(B)/T$ as well as in $\RH(B_2)$ may straightforwardly be related to a jump from a large Fermi surface to a smaller one when decreasing $B$. Remarkably, in both thermopower and Hall effect isotherms the values in the magnetically ordered ground state of \YRS\ (with localized 4$f$ states in the presence of a small Fermi surface) approach the corresponding values of the non-magnetic reference compound LuRh$_2$Si$_2$ \cite{Koehler08,Friedemann08}. In the latter, no 4$f$ states contribute to the (small) Fermi surface either. 

At elevated temperatures, a pronounced logarithmic divergence is revealed in $-S(T)/T$ (cf.\ upper inset of Fig.~\ref{fig:thermopower}(a)). This parallels  observations on $\gamma(T)$ \cite{Trovarelli00,Custers03}. However, in the FL regime ($B>\BN$), the field dependence of the saturation value $\lim_{T\to0}(-S(T)/T)$ extrapolates to a \textit{finite value} at $B=\BN$, where $\gamma(T\to 0)$ \textit{diverges}. These disparate field dependences are at strong variance to the prediction within  theories addressing a SDW QCP \cite{Paul01}.

\begin{figure}
	\begin{center}
		\includegraphics[width=.94\columnwidth]{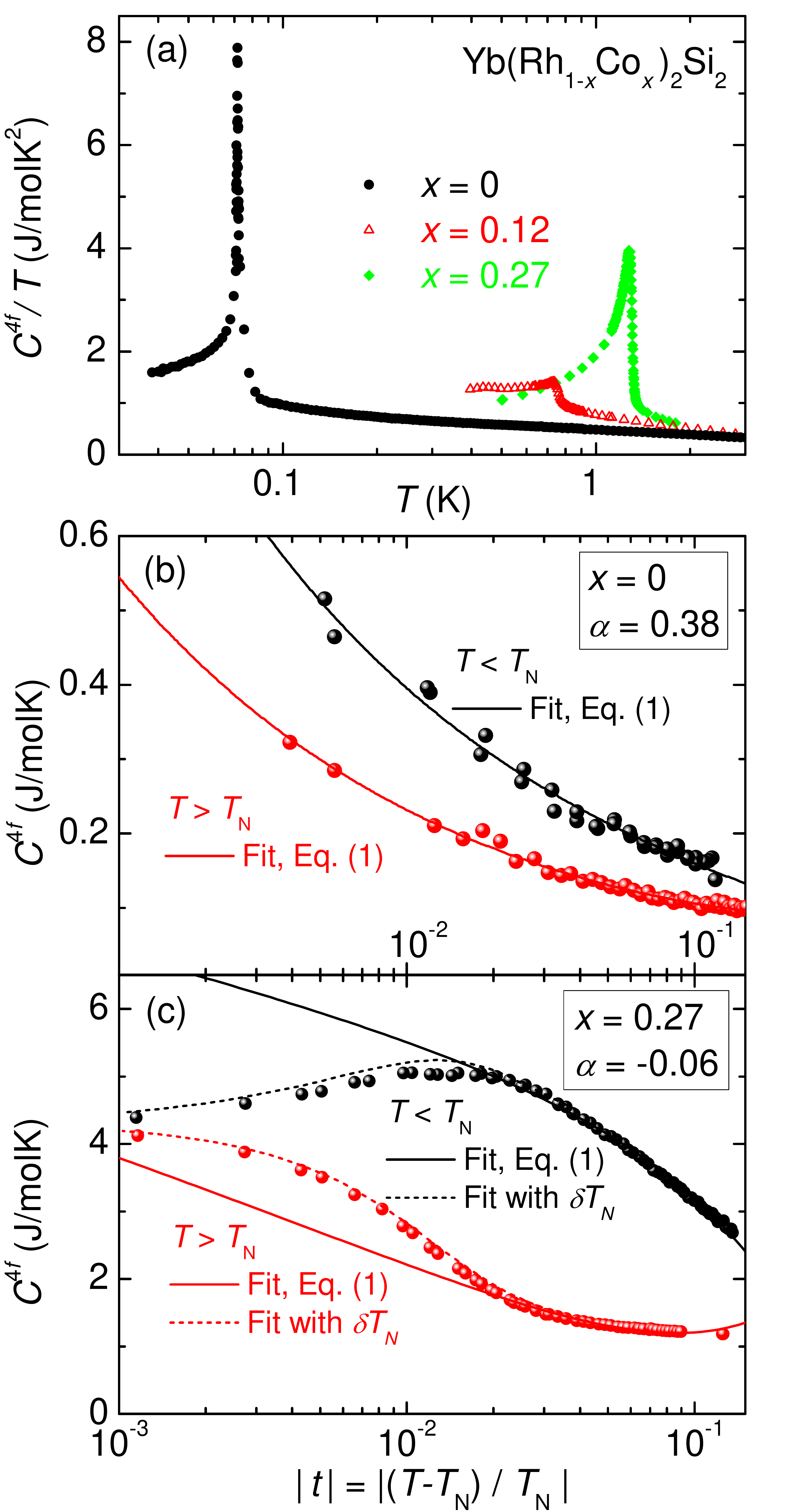}
	\end{center}
 \caption{\label{fig3}(a) $4f$ increment to the specific heat plotted as $C^{4f}/T$ against a logarithmic $T$ scale for three single crystals of the series Yb(Rh$_{1-x}$Co$_x$)$_2$Si$_2$. 
 Analysis of the critical exponent for (b) $x=0$  and (c) $x=0.27$  close to $\TN$ ($|t|\leq 0.1$). The specific heat data (symbols) are fitted using Eq.~\ref{FktFit} (solid lines). Dotted lines in (c) represent the fit including a Gaussian distribution of $\TN$ with $\delta \TN/\TN=6.5\cdot 10^{-3}$ with otherwise identical fit parameters.}
\end{figure}

The proximity of \YRS\ to the unconventional QCP even seems to affect the classical phase transition at \TN\ as here, an unusual scaling exponent is observed. 
In general, universal static scaling dependences are valid for all classical second-order phase transitions and are only determined by the symmetry of the order parameter and the dimensionality of the critical fluctuations \cite{Landau37, Wosnitza07}. 
The high-precision specific-heat measurements on a high-quality \YRS\ single crystal depicted in Fig.~\ref{fig3}(a) allowed for a detailed analysis of the critical fluctuations \cite{Krellner09}. 
The anomaly  due to the onset of magnetic order is clearly visible. 
The critical exponent can be extracted utilizing the standard fit function \cite{Wosnitza07},
\begin{equation}\label{FktFit}
C^{\pm}(t) = \frac{A^{\pm}}{\alpha}|t|^{-\alpha} + b + Dt \textnormal{ ,}
\end{equation}
to describe the critical behavior with the reduced temperature $t=(T - \TN)/\TN$, where $+$($-$) refers to $t>0$ ($t<0$), respectively. 
The background contribution is approximated by a linear $t$ dependence ($b+Dt$) close to $\TN$, while the power law (first term in Eq.~\ref{FktFit}) represents the leading contribution to the singularity in $C^{4f}(t)$. 

The excellent quality of the fit becomes evident in Fig.~\ref{fig3}(b), where the curves are shown below (black lines) and above (red lines) $\TN$, together with the experimental data. For \YRS, the fitting procedure yields a critical exponent of $\alpha=0.38\pm0.03$, which describes the data in the entire temperature range around the phase transition at $\TN=72\pm1$\,mK. Also, no saturation (due to rounding effects of the phase transition) is observed at the lowest $|t|$ values. The exponent derived for stoichiometric \YRS\ strongly deviates from those obtained for any universality class. 
The largest positive value for $\alpha$ is calculated for a 3D-Ising system, so that generally $\alpha\leq 0.11$ \cite{Wosnitza07}. Our surprisingly large critical exponent is supported by thermal-expansion measurements in the vicinity of $\TN$. Thermodynamic relations reveal the same critical exponent for the thermal expansion as for the specific heat \cite{Scheer92}.  Application of the same fitting procedure as for the specific heat gives a critical exponent of the thermal expansion of $\alpha=0.30\pm0.15$, in good agreement with the specific-heat result. It should be noted that minor variations 
of the measured length change around $\TN$ severely complicate the 
critical scaling analysis
and impede a more accurate determination of $\alpha$ \cite{Krellner09}. 
Nevertheless, the discrepancy of the critical exponent found in \YRS\ with known universality classes is robust and likely to originate in the nearby unconventional QCP which appears to substantially influence the spatial fluctuations of the classical order parameter.

\section{Detaching the AF instability and Kondo breakdown by chemical pressure}
\begin{figure}
	\begin{center}
		\includegraphics[trim=.4cm .3cm .7cm .5cm,width=.94\columnwidth]{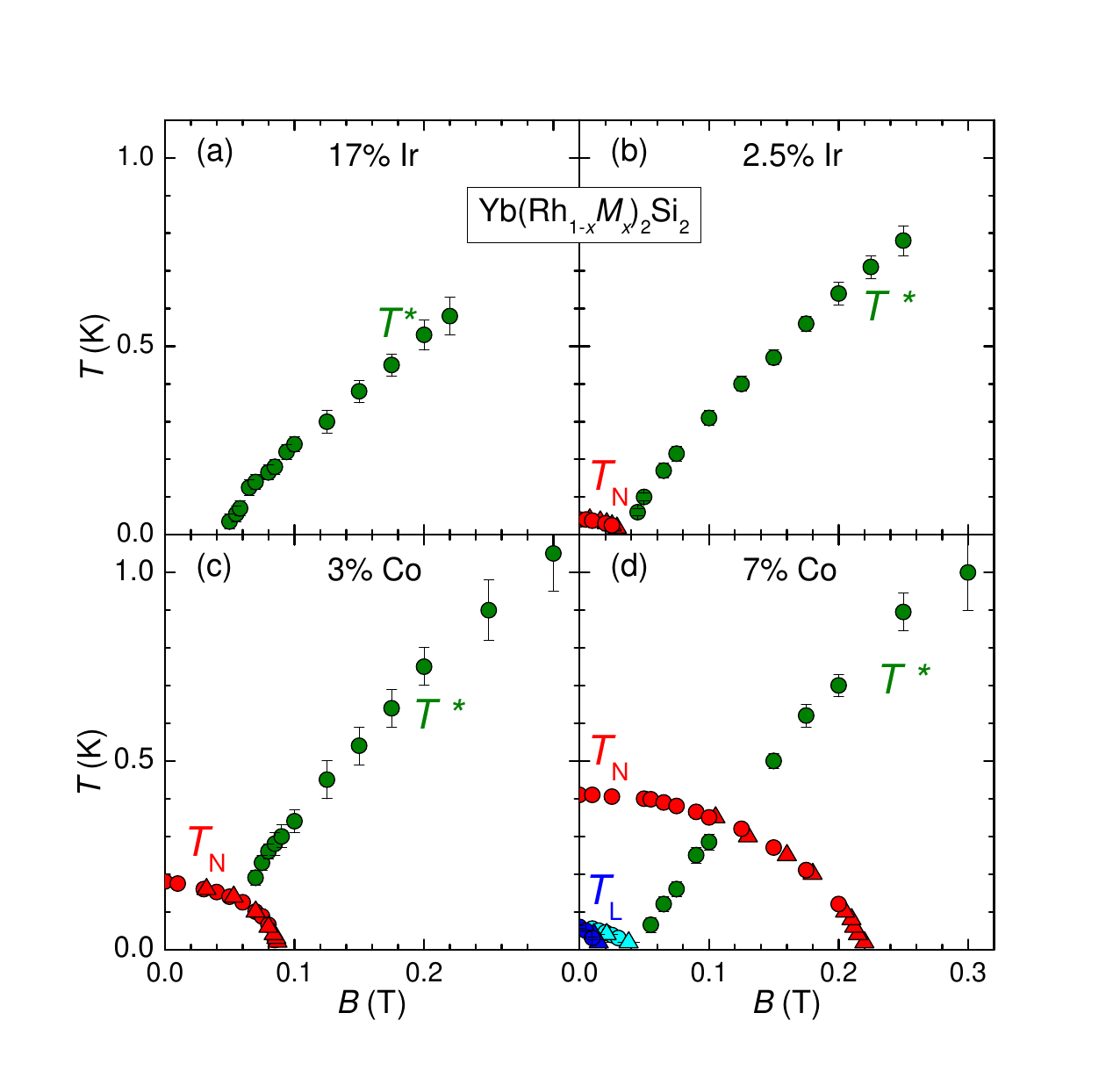}\\
		\includegraphics[trim=.8cm .8cm .8cm .8cm,width=.93\columnwidth]{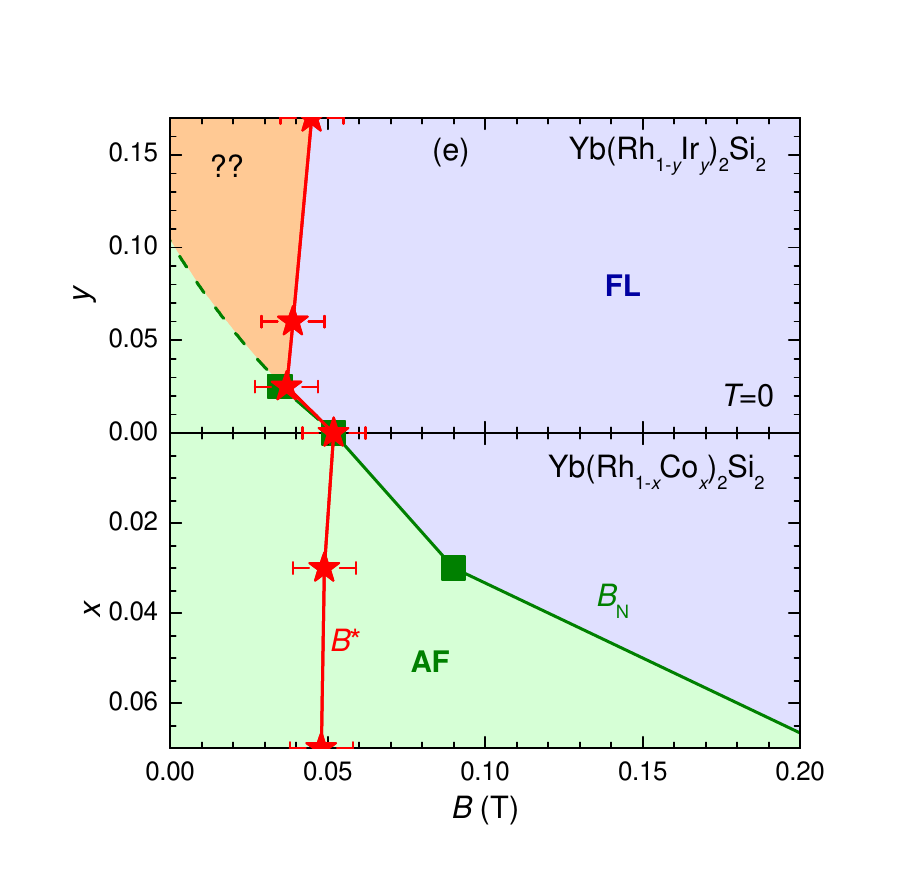}	
	\end{center}
	\caption{ Phase diagrams of 
	Yb(Rh$_{1-x}$\textit{M}$_x$)$_2$Si$_2$
	with 17\% Ir (a), 
	2.5\% Ir (b), 
	3\% Co (c), 
	and 7\% Co (d). 
	The $\Tstar(B)$ line and $\TN(B)$ were deduced from 
	a.c.\ susceptibility $\chi_{\text{ac}}$ measurements 
	where they manifest as a maximum and a cusp in $\chi_{\text{ac}} (T)$. 
	(e) Zero-temperature phase diagram of Yb(Rh$_{1-x}$\textit{M}$_x$)$_2$Si$_2$. 
	The variation of $B^{\star}$ and $B_{\text N}$, the critical fields of the Kondo breakdown energy scale $T^{\star}(B)$ and of the antiferromagnetically ordered phase, respectively, with composition are depicted \cite{Friedemann09}. Green and blue shaded region mark the antiferromagnetically ordered and paramagnetic FL ground state, respectively. For Ir substitution with $y> 2.5$\,\% an new phase emerges which resembles a spin liquid.}
	\label{fig:PD}
\end{figure}

Having discussed before the concurrence of the AF instability and the Kondo breakdown in stoichiometric \YRS\ the question arises whether these two instabilities could be separated by changing an additional control parameter, \textit{i.e.}, volume.
As was found earlier, pressurizing \YRS\ stabilizes magnetism \cite{Mederle02}. 
In a recent study isoelectronic substitution was utilized to generate negative and positive chemical pressure \cite{Friedemann09}. The combination with results obtained under hydrostatic pressure confirmed the dominance of the volume over the disorder effect. For both positive physical and positive chemical pressure, magnetism is strengthened as expected: \TN\ and \BN\ increase with decreasing unit-cell volume.
Ir substitution results in a unit-cell expansion and weakens magnetism. In fact, magnetic order is completely suppressed for Yb(Rh$_{0.83}$Ir$_{0.17}$)$_2$Si$_2$ (Fig.~\ref{fig:PD}(a)) and \TN\ is slightly reduced for Yb(Rh$_{0.975}$Ir$_{0.025}$)$_2$Si$_2$ (Fig.~\ref{fig:PD}  (b)). Substituting Co leads to a unit-cell compression which strengthens magnetism as illustrated for Yb(Rh$_{0.97}$Co$_{0.03}$)$_2$Si$_2$ in Fig.~\ref{fig:PD}(c) and for Yb(Rh$_{0.93}$Co$_{0.07}$)$_2$Si$_2$ in Fig.~\ref{fig:PD}(d). A second transition discovered for \YRS\ under hydrostatic pressure\cite{Mederle02} is also observed for  Yb(Rh$_{0.93}$Co$_{0.07}$)$_2$Si$_2$  at $T_{\text L} = 0.06$\,K. 
The surprising finding is that the Kondo breakdown energy scale is nearly insensitive against volume changes.
This is obvious from Figs.~\ref{fig:PD}(a) to (d) showing the $\Tstar(B)$ line remaining at almost the same position in the $T$-$B$ phase diagram as for pure \YRS\ (cf.\ inset of Fig.~\ref{fig:RHvsB2}(b)). 

The essential result of the chemical pressure study is summarized in the ($T=0$) global phase diagram: Fig.~\ref{fig:PD}(e) displays the critical fields \BN\ and \Bstar\ where $\TN(B)$ and $\Tstar(B)$ extrapolate to zero temperature. 
The decrease of \BN\ for Ir substitution reflects the decline of magnetism under lattice expansion, while the increase for Co substitution reflects the rise of magnetism under lattice compression. Clearly, \Bstar\ remains almost unchanged in the investigated range of substitution. 

The different behaviors of \BN\ and \Bstar\ have strong consequences  leading to two limiting cases. For Co substituted samples we observe a crossing of $\TN(B)$ and $\Tstar(B)$ at finite $T$. This is anticipated for 3\% Co content (Fig.~\ref{fig:PD}(c) and Ref. \cite{Friedemann10d}). For  Co content $x\geq 7\%$ the signatures of the $\Tstar(B)$-line are clearly observed within the magnetically ordered phase. This finding is confirmed by a study on pure \YRS\ under physical pressure \cite{Tokiwa09}.
Also no relation between the second transition at \TL\ and the $\Tstar(B)$ line is evident as only for Yb(Rh$_{0.93}$Co$_{0.07}$)$_2$Si$_2$ \TL\ vanishes in proximity to \Bstar\ line rendering this an accidental coincidence.
With \Bstar\ assigned to the Kondo breakdown, we infer from $\BN > \Bstar$ that the magnetism extends into the regime where the Kondo effect is operating. Hence, the magnetic instability is expected to obey the signatures of an SDW QCP. This is in accordance with the observed power-law dependence of $\TN(B)\propto (\BN-B)^n$, yielding an exponent $n=0.65$ in surprisingly good agreement with $n=2/3$ predicted for 3D SDW fluctuations \cite{Friedemann09}. 

In addition, the critical exponent $\alpha$ of the specific heat recovers a conventional value for \YCxS. Fig.~\ref{fig3}(a) includes specific heat data on two selected Co-concentrations ($x=0.12$ and 0.27). Already the shape of the anomalies indicates a change of the critical fluctuations with increasing $x$. Going from $x=0$ to $x=0.12$, the phase transition anomaly at \TN\ changes drastically, \textit{i.e.}, from 
a very sharp $\lambda$-type peak to 
a rather broad mean-field-like anomaly \sven{detected for concentrations $x\geq 0.07$ (not shown)} \cite{Steppke2010}, which prevents the extraction of critical fluctuations. For $x=0.27$ one again observes a pronounced $\lambda$-type anomaly which allows to analyze the data in terms of critical fluctuations.

In Fig.~\ref{fig3}(c), we present the analysis for  $x=0.27$ sample. Here, $\TN=1.298 
$\,K exceeds \TN\ of pure \YRS\ by one order of magnitude.
The best fit  for $x=0.27$ reveals a negative critical exponent $\alpha=-0.06\pm0.10$, which can already be inferred from the non-diverging $t$-dependence in Fig.~\ref{fig3}(c) (for comparison see Fig.~1 in Ref. \cite{Wosnitza07}). To 
fit the data points at $|t|\leq 0.01$ we have to allow for a Gaussian distribution of $\TN$ with $\delta \TN=8.4$\,mK to account for the rounding effects due to the high substitution level. This smearing of $\TN$ leads to a relatively small temperature range, $0.01\leq |t|\leq 0.1$, which determines the critical exponent, 
impeding a more accurate determination of $\alpha$. A similar measurement and analysis of the critical behavior was also carried out for $x=0.38$, yielding again a negative critical exponent $\alpha=-0.12\pm0.10$ \cite{Krellner10}.

Comparing the critical exponents for the different materials, it is obvious that there is a drastic change on going from $\alpha=0.38(3)$ for $x=0$ to $\alpha=-0.06(10)$ for $x=0.27$. The latter value can be explained in terms of the classical universality classes in the Landau theory of phase transitions, for which $-0.133(5)\leq \alpha \leq +0.110(1)$ generally holds true.  However, the accuracy of the determined exponent for the  samples with substitution is not sufficient to distinguish between the two applicable symmetry classes \cite{Wosnitza07}, namely the 3D-Heisenberg model $[\alpha_{3D,H}=-0.133(5)]$ or the 3D-XY model $[\alpha_{3D,XY}=-0.015(1)]$. 

The evolution of the phase transition anomaly at \TN\ in $C^{4f}(T)/T$  can be understood in the frame of Doniach's phase diagram  applied to Yb compounds \cite{Gegenwart08}. At large Co concentrations $x$ (small average unit-cell volume) where the RKKY interaction predominates, classical ordering between local 4$f$-derived magnetic moments occurs. Upon lowering $x$ the Kondo effect comes into play. This leads to mean-field type itinerant AF (SDW) order in the renormalized electron fluid. However, as $x\to 0$ the Kondo effect breaks down and the local moments reappear -- fully unexpected within the Doniach phase diagram. Because of the small value of \TN, the nearby quantum critical electronic fluctuations  associated with this Kondo breakdown QCP add to the classical critical AF order parameter fluctuations and very likely contribute to the unusually large critical exponent found for pure \YRS.


Furthermore, our result of a conventional critical exponent for chemically pressurized \YRS\ is in contradiction to the quantum-tricritical-point scenario \cite{Misawa09}. Here, Misawa \textit{et al.} suggest a stabilized tricritical point at finite temperatures in pressurized \YRS. As the theoretical value at a classical tricritical point is $\alpha=0.5$, one would expect that the critical exponent becomes even larger for \YRS\ under chemical pressure, just opposite to our observation.

The findings on Co substituted \YRS\ are resembled by two other heavy-fermion systems: Both CeIn$_3$ and CeRh$_{1-x}$Co$_x$In$_5$ display a Fermi surface reconstruction inside the antiferromagnetically ordered state \cite{Harrison07,Goh08}. For the case of CeIn$_3$ this suggests a transition from a large Fermi surface with the $f$-electrons incorporated to a small Fermi surface with the $f$ electrons decoupled from the conduction sea. For CeRh$_{1-x}$Co$_x$In$_5$ the Fermi surface reconstruction appears to be accompanied by a change in the magnetic structure. 

We now turn to the samples with Ir substitution, at negative chemical pressure where $\BN<\Bstar$. Thus, for instance, for 6\% Ir substituted \YRS\ $\BN \approx 15$\,mT is well below the Kondo-breakdown critical field $\Bstar\approx 40$\,mT. Consequently, there exists a finite field range at $T=0$ (indicated by question marks in Fig.~\ref{fig:PD}(e)) in which the magnetic moments are neither ordered nor screened by the Kondo effect. The electrical resistivity exhibits pronounced NFL behavior in this regime which was tentatively ascribed to a metallic spin-liquid phase, hitherto unknown in a Kondo-lattice system \cite{Friedemann09}. Similar observations were made with other lattice expanded \YRS\ samples, substituted either with La on the Yb site or Ge on the Si site\cite{Weickert06,Custers10}.


The most prominent point in the global phase diagram (Fig.~\ref{fig:PD}(e)) is taken by stoichiometric \YRS. Here, the AF instability and the Kondo breakdown concur, $\BN = \Bstar$. Within experimental resolution this concurrence extends towards small Ir content
 as obvious from 
 Fig.~\ref{fig:PD}(b) 
\cite{Friedemann10c}. 
For Yb(Rh$_{0.975}$Ir$_{0.025}$)$_2$Si$_2$  the critical field of the Néel phase is slightly reduced compared to pure \YRS. Apparently, the critical field of the $\Tstar(B)$-line seems to be reduced by almost the same amount. Consequently, \BN\ and \Bstar\ concur within experimental accuracy for this slightly Ir substituted sample.
Such a concurrence between the AF and electronic instabilities in a finite parameter range of the global phase diagram for a Kondo lattice has been predicted theoretically \cite{Si10,Si06}


\section{Conclusion and Perspective}
We have provided compelling evidence for a breakdown of the Kondo effect overlapping with the field induced AF QCP in pure \YRS. This is found to be accompanied by a reconstruction of the Fermi surface and highly unusual scaling behavior. Upon varying the average unit-cell volume of \YRS\ by partial isoelectronic substitution these very different instabilities can be separated from each other. Volume compression (Co substitution) furnishes an intersection of the Kondo-breakdown crossover line $\Tstar(B)$ with the magnetic phase boundary $\TN(B)$. Here, the AF QCP ($\TN\to0$) is of  conventional 3D SDW type. Future studies on moderately Co substituted samples can establish the existence of a  quantum phase transition between local and itinerant AF order. 

When the unit cell of \YRS\ is moderately expanded, a novel low-temperature NFL phase, presumably of spin-liquid type, develops in a wide field range. Whether such a phase can in fact arise within the Kondo lattice model is an intriguing open theoretical question. Finally, \YRS\ with sufficiently expanded volume (\textit{e.g.} containing more than 10\% Ir) should be investigated to unravel the nature of the Kondo breakdown QCP without any interfering magnetism.
 \sven{At small Ir concentrations a finite range seems to exist where the magnetic and electronic instabilities concur. Specific heat measurements (\textit{e.g.} on samples with $\approx 2.5$\% Ir) may reveal whether the unusual scaling behavior is linked to this concurrence.}
 
\acknowledgements
We greatly acknowledge valuable discussions with  P.~Coleman, G.~Donath, P.~Gegenwart, C.~Klingner, M.~Nicklas, N.~Oeschler, S.~Paschen and G.~Zwicknagl. This works was supported by the DFG Research Group 960 ``Quantum Phase Transitions'' as well as by the NSF Grant No.~DMR-1006985 and the Robert Welch Foundation Grant No.~C-1411.


\end{document}